\documentclass{article}

\usepackage{arxiv}

\usepackage[utf8]{inputenc}
\usepackage[T1]{fontenc}
\usepackage{hyperref}
\usepackage{url}
\usepackage{booktabs}
\usepackage{amsfonts}
\usepackage{nicefrac}
\usepackage{microtype}
\usepackage{graphicx}
\usepackage{float}
\usepackage[breakable]{tcolorbox}
\tcbset{
  casebox/.style={colback=gray!5, colframe=gray!60, fonttitle=\bfseries\small, boxrule=0.5pt, left=4pt, right=4pt, top=3pt, bottom=3pt, breakable},
}
\graphicspath{{./figures/}}

\title{SkillForge: Forging Domain-Specific, Self-Evolving Agent Skills in Cloud Technical Support}

\author{
  Xingyan Liu$^*$\quad
  Xiyue Luo\quad
  Linyu Li\quad
  Ganghong Huang\quad
  Jianfeng Liu$^*$\quad
  Honglin Qiao$^*$ \\
  Alibaba Cloud Computing, Alibaba Group \\
  \texttt{\{wxw, luoxiyue.lxy, lly455313, huangganghong.hgh, jiawei.ljf, kenny.qhl\}@alibaba-inc.com}
}

\date{}

\begin{document}
\maketitle

%% --- 3. Abstract ---
\begin{abstract}
Deploying LLM-powered agents in enterprise scenarios such as cloud technical support demands high-quality, domain-specific skills. However, existing skill creators lack domain grounding, producing skills poorly aligned with real-world task requirements. Moreover, once deployed, there is no systematic mechanism to trace execution failures back to skill deficiencies and drive targeted refinements, leaving skill quality stagnant despite accumulating operational evidence.
We introduce SkillForge, a self-evolving framework that closes an end-to-end creation--evaluation--refinement loop. To produce well-aligned initial skills, a Domain-Contextualized Skill Creator grounds skill synthesis in knowledge bases and historical support tickets. To enable continuous self-optimization, a three-stage pipeline --- Failure Analyzer, Skill Diagnostician, and Skill Optimizer --- automatically diagnoses execution failures in batch, pinpoints the underlying skill deficiencies, and rewrites the skill to eliminate them. This cycle runs iteratively, allowing skills to self-improve with every round of deployment feedback.
Evaluated on five real-world cloud support scenarios spanning 1,883 tickets and 3,737 tasks, experiments show that: (1) the Domain-Contextualized Skill Creator produces substantially better initial skills than the generic skill creator, as measured by consistency with expert-authored reference responses from historical tickets; and (2) the self-evolution loop progressively improves skill quality from diverse starting points (including expert-authored, domain-created, and generic skills) across successive rounds, demonstrating that automated evolution can surpass manually curated expert knowledge.
\end{abstract}

\textbf{Keywords:} LLM Agents, Agent Skills, Self-Evolution, Cloud Technical Support

\footnotetext{$^*$Corresponding authors. This paper has been accepted at ACM SIGIR 2026 Industry Track. This version includes extended appendices and a related work section.}

%% --- Main Content ---

\section{Introduction}

Enterprise technical support demands reliable procedural knowledge and organizational context beyond raw language modeling~\cite{wang2024llmagentsurvey}, and LLM-based agents are increasingly deployed in cloud operations---from automated incident root cause analysis to domain-specific diagnosis~\cite{chen2024rcacopilot,wang2023rcagent,zhou2024dbot}. The Agent Skills concept formalizes this specialization: a portable, file-based package that encapsulates procedures, resources, and tool-use guidance so agents can be versioned and composed like software modules~\cite{anthropicAgentSkills2025}. Several foundations now underpin skill-equipped agents: tool-augmented architectures enable robust API invocation and reasoning over results~\cite{react2023,toolformer2023,qin2024toolllm}; iterative self-improvement methods let LLMs refine outputs through feedback loops~\cite{madaan2023selfrefine,shinn2023reflexion,zhao2024expel}; and automatic instruction optimization techniques can discover stronger agent directives~\cite{zhou2023ape,yang2023opro,yuksekgonul2024textgrad}. However, these approaches operate at the prompt or single-task level rather than managing reusable skill artifacts that can be systematically created with domain knowledge and continuously improved from deployment feedback.

Producing and maintaining such high-quality skills in enterprise settings remains difficult due to two key challenges: (1) generic skill creators lack domain grounding, producing poorly aligned initial skills; and (2) no systematic mechanism exists to trace execution failures back to skill deficiencies and drive targeted refinements. SkillsBench, a recent benchmark of 86 tasks across 11 domains, confirms that curated skills improve success rates by 16.2pp on average while self-generated skills provide no benefit~\cite{skillsbench2026}. Concurrent work on continuous improvement of LLM-based customer support relies on human annotation as the feedback signal~\cite{zhao2025agentloop}; SkillForge, by contrast, directly targets the skill artifact itself---automating failure analysis and skill rewriting without requiring per-deployment human supervision. We present SkillForge, an end-to-end creation--evaluation--refinement loop that addresses both challenges: a Domain-Contextualized Skill Creator for high-quality initialization, and an automated diagnosis-optimization pipeline for continuous self-evolution. We validate the framework on five real-world cloud technical support scenarios.

\section{The Self-Evolving Skill Framework}

\subsection{Application Context}
This work focuses on customer service AI agents deployed in cloud technical support. The agent's primary output is a customer-facing reply, which is presented to human support engineers for review. In production, the human agent decides whether to adopt the AI-generated response (with or without minor edits) before sending it to the customer. During the problem-solving process, the AI agent may invoke various tools---including knowledge retrieval, diagnostic APIs, and resource queries---to assist in formulating the response. This human-in-the-loop design enables the ``consistency with expert reference'' evaluation paradigm used throughout this paper: we compare the agent's output against the actual solutions provided by human experts in historical tickets.

\subsection{Anatomy of an Agent Skill}
\label{sec:anatomy}
A standard Agent Skill is a self-contained package comprising \texttt{SKILL.md} (instructions and workflow logic), \texttt{scripts/} (executable code), and \texttt{references/} (domain documentation)~\cite{anthropicAgentSkills2025}. In enterprise environments, however, allowing agents to execute arbitrary scripts poses security and stability risks. We therefore adopt a constrained definition: the skill excludes \texttt{scripts/} and relies exclusively on pre-defined, verified system tools. We add \texttt{references/tools.json} to store schemas for these verified tools.

Correspondingly, all skill-bearing agents---including both business agents handling customer tickets and meta-agents (Diagnostician, Optimizer) managing skill evolution---interact with skill assets (\texttt{SKILL.md}, \texttt{references/}) exclusively through a \textbf{Virtual File System (VFS)} rather than executing arbitrary code. This design is motivated by several considerations: (1) the majority of high-frequency customer service tasks can be effectively addressed through instruction/knowledge injection and pre-defined tools without requiring dynamic code execution; (2) eliminating executable scripts significantly improves runtime stability in production; and (3) constraining the action space to text-based operations simplifies failure diagnosis and optimization. More complex scenarios requiring real-time sandbox script execution are left to future work.

\subsection{Framework Overview}
\begin{figure}[H]
  \centering
  \includegraphics[width=\linewidth]{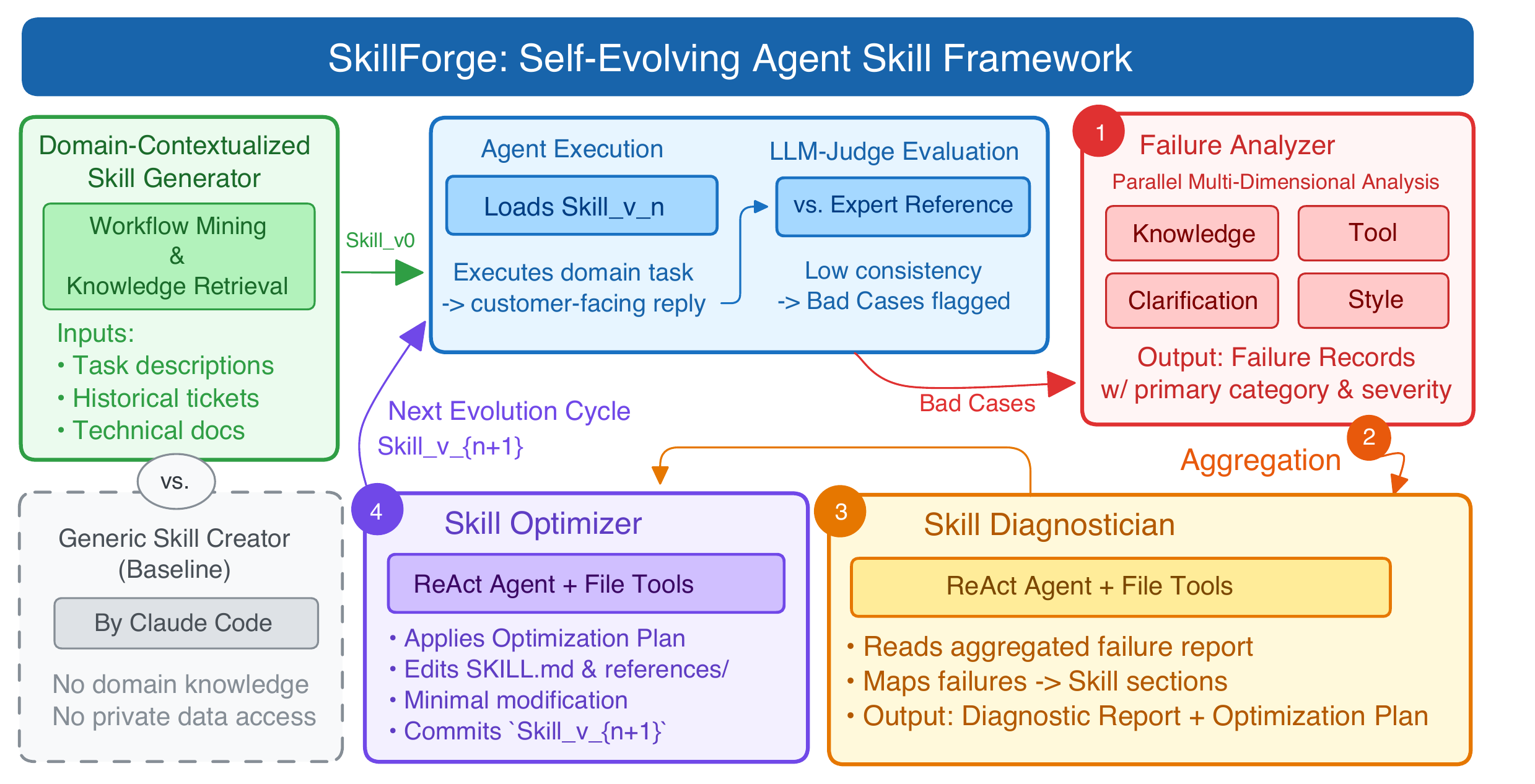}
  \caption{Overview of the SkillForge self-evolving skill framework. A Domain-Contextualized Skill Creator mines historical tickets and domain knowledge to produce an initial \texttt{Skill\_v0}. An iterative loop of Failure Analysis, Root Cause Diagnosis, and Skill Optimization then continuously refines the skill based on accumulated bad cases from agent execution.}
  \label{fig:framework}
\end{figure}

Our framework establishes a continuous improvement cycle inspired by recent work on self-evolving agents~\cite{zhuge2024symbolic,shinn2023reflexion,memskill2026,hu2024adas}, designed as a five-phase pipeline to ensure robustness and traceability (Figure~\ref{fig:framework}):
\begin{enumerate}
    \item \textbf{Initialization:} The \texttt{Domain-Contextualized Skill Creator} generates \texttt{Skill\_v0} based on domain data.
    \item \textbf{Execution \& Monitoring:} The agent executes tasks using \texttt{Skill\_v\_n}. Discrepancies between agent execution and reference behaviors are flagged as \textbf{Bad Cases}.
    \item \textbf{Phase 1: Multi-Dimensional Failure Analysis:} The \textbf{Failure Analyzer} processes each bad case in parallel across four dimensions (Knowledge, Tool, Clarification, Style) to produce \texttt{Structured Failure Records}.
    \item \textbf{Phase 2: Aggregation:} Individual failure records are aggregated by category to identify systemic patterns and select representative cases.
    \item \textbf{Phase 3: Diagnosis:} The \textbf{Skill Diagnostician} analyzes the aggregated data to map failures to specific sections of the skill definition, producing a \texttt{Diagnostic Report} and \texttt{Optimization Plan}.
    \item \textbf{Phase 4: Optimization:} The \textbf{Skill Optimizer} references both the \texttt{Diagnostic Report} and the \texttt{Optimization Plan} to apply targeted edits to \texttt{Skill\_v\_n} via the \textbf{Virtual File System}, producing \texttt{Skill\_v\_n+1}.
\end{enumerate}

\subsection{Domain-Contextualized Skill Creator}
The creator addresses the cold-start problem by generating a robust initial skill. Due to enterprise data privacy requirements, we use an internal LLM (Qwen3-Max) to process proprietary domain data.
\begin{itemize}
    \item \textbf{Input:} Task descriptions, historical ticket datasets, and technical documentation.
    \item \textbf{Process:}
    \begin{itemize}
        \item \textbf{Workflow Mining:} Extracts typical solution patterns (e.g., Clarify $\rightarrow$ Diagnose $\rightarrow$ Resolve) from historical ticket dialogues and operation sequences.
        \item \textbf{Tool Mining:} Identifies high-frequency tools used by human experts in historical tickets and extracts their schemas for inclusion in the skill definition.
        \item \textbf{Knowledge Extraction:} Searches internal documentation and knowledge bases (or collects references cited in historical tickets) to extract domain-specific information required by the skill.
        \item \textbf{Skill Synthesis:} Fills a pre-defined cloud service Skill template with the mined workflows, tools, and extracted knowledge.
    \end{itemize}
    \item \textbf{Output:} A structured, domain-rich initial skill (\texttt{Skill\_v0}). \texttt{SKILL.md} is organized into five sections---\textit{Background Knowledge} (concepts most commonly misunderstood by customers), \textit{Case-Type Triage} (a decision tree routing each incoming request to the right handler), \textit{Per-Case-Type Handling} (typically 4--8 case types within the skill, each with a branching workflow, tool guidance, failure-cause/resolution pairs, and an escalation fallback), \textit{FAQ} (long-tail issues with frequency below 5\% of the scenario's tickets), and \textit{Reference Index}. Synthesis enforces that every procedure traces to a mined resolution path, workflows encode conditional branching rather than linear sequences, and detailed content is offloaded to reference files when \texttt{SKILL.md} exceeds 500 lines.
\end{itemize}
See Appendix~\ref{sec:appendix_creator} for details on each mining stage.

\subsection{Agent Execution and Performance Monitoring}

\textbf{Agent Execution.} The agent execution pipeline consists of: (1) identifying the appropriate skill for the current scenario; (2) dynamically loading the skill's \texttt{SKILL.md} and \texttt{references} into context (treating the skill as a ``Meta-tool'' rather than polluting the global system prompt); (3) executing the task using the loaded instructions and available tools; and (4) returning the generated response.

\textbf{Performance Evaluation.} The agent loads the skill to handle tasks. The final response is checked for consistency against a reference response (using an LLM-judge). Low consistency indicates a failure, triggering the bad-case analysis pipeline. In this paper we use offline replays of historical tickets; in deployment, online failed interactions (e.g., low-consistency outputs or rejections by human agents) feed the same pipeline. 
See Appendix~\ref{sec:appendix_agent_loop} for implementation details of the agent execution pipeline.

\subsection{Automated Diagnosis}

\subsubsection{Multi-Dimensional Failure Analysis (Failure Analyzer)}
Instead of a single classification, the Failure Analyzer performs a parallel analysis of each Bad Case across four distinct dimensions, drawing on structured reasoning approaches~\cite{wei2022cot,yao2023tot} to ensure comprehensive attribution:
\begin{itemize}
    \item \textbf{Knowledge Analysis:} Checks for missing, incorrect, or contradictory domain knowledge.
    \item \textbf{Tool Analysis:} Evaluates tool calls for missing invocations, wrong parameters, or misinterpretation of results.
    \item \textbf{Clarification Analysis:} Assesses whether the agent over-asked, under-asked, or asked irrelevant questions.
    \item \textbf{Style Analysis:} Reviews the response tone, ensuring it is not robotic, overly verbose, or cold.
\end{itemize}
For each bad case, results from the four dimensions are combined by deterministic rules into a single verdict: a case is marked \textit{fail} when any dimension scores high or at least two score medium; ties on the primary category are broken by a fixed priority knowledge $>$ tool $>$ clarification $>$ style. 
The output is a structured JSON record containing \texttt{failure\_categories}, \texttt{overall\_severity}, \texttt{primary\_category}, and per-dimension diagnostic hints (see Appendix~\ref{sec:appendix_fa}).

\subsubsection{Root Cause Diagnosis (Skill Diagnostician)}
The diagnosis phase bridges the gap between individual failures and skill definition defects:
\begin{itemize}
    \item \textbf{Batch Aggregation (Phase 2):} Per-case failure records from the Analyzer are aggregated across the batch by category to compute severity distributions and issue-type frequencies (e.g., \texttt{knowledge:missing} occurred 8 times) and to select top-$k$ representative cases per category by severity and diversity.
    \item \textbf{Diagnosis Agent (Phase 3):} A ReAct-based~\cite{react2023} \textbf{Skill Diagnostician} agent is invoked. It reads the aggregated report and the current \texttt{SKILL.md} (via a Virtual File System).
    \item \textbf{Process:} The agent operates in four stages---\textit{Skill Understanding} (read and summarize \texttt{SKILL.md} structure), \textit{Evidence Collection} (parse aggregated FA results), \textit{Root Cause Attribution} (map issues to specific \texttt{SKILL.md} locations and classify defect types as \textit{missing}, \textit{insufficient}, or \textit{incorrect}), and \textit{Plan Generation}. Typical attributions include \texttt{knowledge:missing} $\rightarrow$ scenario-section knowledge gap, \texttt{tool:wrong\_params} $\rightarrow$ insufficient parameter documentation, and \texttt{clarification:over\_asking} $\rightarrow$ over-broad gathering rules.
    \item \textbf{Output:} A structured \texttt{Diagnostic Report} and a machine-parsable \texttt{Optimization Plan} in which each recommendation specifies the modification location, content change, whether examples or knowledge search are needed, and expected impact.
\end{itemize}
See Appendix~\ref{sec:appendix_diag_opt} for the full diagnostic workflow and optimization execution details.

\subsection{Automated Optimization (Skill Optimizer)}
The Skill Optimizer applies the optimization plan through the VFS described in Section~\ref{sec:anatomy}.
\begin{itemize}
    \item \textbf{Knowledge Augmentation:} The optimizer performs knowledge searches to fill gaps identified in the diagnostic report and places results by type: \emph{high-frequency, stable} knowledge is added to the relevant skill section; long-tail items lacking authoritative documentation are appended to the FAQ; when official documentation is available, the skill cites the source link rather than embedding the content, leaving full retrieval to the Agent at runtime (we revisit this layered policy in Section~\ref{sec:hallucination}).
    \item \textbf{Process:} The agent executes the Optimization Plan under three core principles: \textit{Minimal Modification} (only change what is necessary to address diagnosed issues), \textit{Do No Harm} (preserve existing correct behaviors via additive changes; never delete working content), and \textit{Evidence-Based} (every change traces to specific failure-analysis evidence). New content is placed by type---background knowledge into dedicated knowledge sections, tool-call rules embedded in the relevant workflow steps, style guidelines near response templates---and detailed content is offloaded to reference files when \texttt{SKILL.md} grows beyond the same 500-line threshold used by the Creator.
    \item \textbf{Versioning:} The modified VFS state is committed as \texttt{Skill\_v\_n+1}, ensuring full traceability.
\end{itemize}

\section{Experiments and Results}

\subsection{Experimental Setup}

Our evaluation methodology follows established agent benchmarking practices~\cite{liu2023agentbench,kapoor2024agents}.

\subsubsection{Scenarios and Dataset}
We evaluate SkillForge on \textbf{five} representative cloud technical support scenarios from a major cloud provider, as summarized in Table~\ref{tab:scenarios}. These scenarios were selected from high-frequency areas already deployed in production---each with established predefined workflows, accumulated human-expert experience, and mature tool infrastructure---to enable direct comparison with the production legacy system. Each scenario corresponds to a top-level cloud product (e.g., DNS) and contains multiple deployed skills, each targeting one major sub-scenario within that product (e.g., ``DNS resolution failure''); skill labels are taken from online production tags originally obtained via ticket clustering and subsequent manual refinement. Each skill in turn covers a handful of case types---the sub-problems automatically distinguished by the Skill Creator.

\begin{table}[H]
\centering
\caption{Evaluation Scenarios and Dataset Statistics.}
\label{tab:scenarios}
\small
\begin{tabular}{l ccccc c}
\toprule
& \textbf{S1} & \textbf{S2} & \textbf{S3} & \textbf{S4} & \textbf{S5} & \textbf{Total} \\
\midrule
\textbf{Scenario} & Account & Domain & DNS & OSS & ECS & -- \\
\textbf{\#Tickets} & 389 & 527 & 256 & 385 & 326 & \textbf{1883} \\
\textbf{\#Tasks} & 706 & 1061 & 572 & 730 & 668 & \textbf{3737} \\
\bottomrule
\end{tabular}
\end{table}

All tickets are real-world, anonymized production tickets. We define a \textbf{task} as a single-turn dialogue within a ticket that can advance ticket resolution---the input is the ticket's message history (including the current user query), and the agent returns a customer-facing reply. A single ticket may involve multiple tasks. Within each scenario, tickets are sorted chronologically and partitioned into four equal-sized splits (25\% each). The first three splits form the \textit{development set} and are fed sequentially to the three evolution rounds (v1, v2, v3) to simulate bad cases arriving in production batches over time; the fourth split is held out as the \textit{evaluation set}, never seen during evolution, and is used to assess the generalizability of the optimizations. Bad cases are identified through automated LLM-judge filtering.

\subsubsection{Evaluation Metrics}

We employ an LLM-judged metric---\textbf{Consistency Rate (CR)}---for skill quality evaluation, comparing the agent's response against expert reference responses from historical tickets. The LLM-judge classifies each response into one of three categories:
\begin{itemize}
    \item \textbf{Consistent:} The clarification questions and solution are aligned with the reference; minor phrasing differences do not affect problem resolution.
    \item \textbf{Partially Consistent:} The response overlaps with the reference without contradiction, but may miss some details.
    \item \textbf{Inconsistent:} The response lacks critical clarifications, misses core solution elements, or conflicts with the reference.
\end{itemize}
The judge first extracts the \textit{core action} from each reference (filtering boilerplate such as greetings or closing remarks) and treats the reference as \textit{one} acceptable solution rather than the unique correct answer---an alternative response is also Consistent if it correctly resolves the customer's problem in line with the ticket's final resolution. 
We report two variants: \textbf{Strict CR} (proportion classified as Consistent) and \textbf{Lenient CR} (Consistent or Partially Consistent). We validated the LLM-judge against human annotations on a sample subset: each sample was independently labeled by two domain experts, and the judge achieved over 90\% agreement with the human consensus, confirming its reliability for automated evaluation (see Appendix~\ref{sec:appendix_llm_judge} for evaluation criteria and output schema).

\subsubsection{Baselines and Variants}
\begin{itemize}
    \item \texttt{S\_generic}: Skills generated by a generic skill creator (Claude Code with Claude-Sonnet-4.5) augmented with domain-specific tool schemas mined from historical tickets, but without access to domain knowledge or historical ticket content due to enterprise data privacy constraints.
    \item \texttt{S\_domain}: Initial output of our Domain-Contextualized Skill Creator.
    \item \texttt{S\_manual}: Manually authored initial skill by human domain experts, encoding expert knowledge and best practices.
\end{itemize}
To evaluate the generality of our self-evolution loop, we use \texttt{S\_manual}, \texttt{S\_domain}, and \texttt{S\_generic} as different starting points for iterative evolution (v1/v2/v3 denote successive evolution cycles).

\subsubsection{Implementation Details}

All experiments use the latest version of Qwen3-Max as the backbone LLM. Each offline evaluation is repeated 3 times; we report the mean. Bad cases are dynamically identified by the LLM-judge after the agent executes each task with the current skill version.

We structure our evaluation around two research questions: \textbf{RQ1}---Does domain-contextualized skill creation produce higher-quality initial skills than a generic creator? \textbf{RQ2}---Can the self-evolution loop progressively improve skill quality from diverse starting points?

% ------------------
\subsection{RQ1: Efficacy of Domain-Contextualized Skill Creator}

We compare the initial skill quality of our domain-aware generator (\texttt{S\_domain}) against the generic baseline. Table~\ref{tab:rq1} presents the results.

\begin{table}[H]
\centering
\caption{RQ1: Comparison of initial skill quality across scenarios. Format: Strict CR (Lenient CR). Best Strict CR in \textbf{bold}.}
\label{tab:rq1}
\footnotesize
\begin{tabular}{l ccccc}
\toprule
\textbf{Method} & \textbf{S1} & \textbf{S2} & \textbf{S3} & \textbf{S4} & \textbf{S5} \\
\midrule
\texttt{S\_generic} & 58.0 (61.4) & 57.9 (64.4) & 63.0 (69.6) & 59.1 (66.3) & 43.4 (55.3) \\
\texttt{S\_domain}  & \textbf{64.0} (65.7) & \textbf{60.5} (68.8) & \textbf{65.4} (70.6) & \textbf{62.5} (72.7) & \textbf{50.6} (57.2) \\
\bottomrule
\end{tabular}
\end{table}

\textbf{Analysis.} \texttt{S\_domain} outperforms \texttt{S\_generic} across all scenarios, with average gains of \textbf{+4.3pp} Strict CR and \textbf{+3.6pp} Lenient CR. The improvement is consistent across all five scenarios, with the largest Strict CR gain in S5 (+7.20pp). Since \texttt{S\_generic} already includes mined tool schemas, the gap confirms that domain-specific workflow knowledge and knowledge extraction provide additional value beyond tools alone.

% ------------------
\subsection{RQ2: Effectiveness of the Self-Evolution Loop}

To evaluate the generality of the self-evolution mechanism, we apply it from three distinct starting points---\texttt{S\_manual}, \texttt{S\_domain}, and \texttt{S\_generic}---and track the improvement over three evolution cycles. Table~\ref{tab:rq2_iteration} reports the relative gain ($\Delta$) in Strict CR and Lenient CR on the held-out evaluation set.

\begin{table}[H]
\centering
\caption{RQ2: Consistency Rate improvement ($\Delta$) across self-evolution iterations from different starting points on held-out evaluation set. Format: Strict CR (Lenient CR).}
\label{tab:rq2_iteration}
\small
\begin{tabular}{l ccc}
\toprule
\textbf{Iteration} & \texttt{S\_manual} & \texttt{S\_domain} & \texttt{S\_generic} \\
\midrule
v1 ($\Delta$) & +4.09 (+5.39) & +2.36 (+0.24) & +7.70 (+1.60) \\
v2 ($\Delta$) & +9.64 (+9.94) & +7.31 (+4.13) & +8.40 (+4.60) \\
v3 ($\Delta$) & \textbf{+10.99} (+12.21) & +9.23 (+8.00) & +11.60 (+4.90) \\
\bottomrule
\end{tabular}
\end{table}

\textbf{Analysis.} Several key observations emerge:
\begin{itemize}
    \item \textbf{Universal Improvement:} All three starting points benefit from the self-evolution loop, with Strict CR gains of +10.99, +9.23, and +11.60 after three iterations respectively. This confirms that the framework is effective regardless of the initial skill quality.
    \item \textbf{Monotonic Progress:} Cumulative gains increase at every iteration from each starting point, confirming that successive cycles add positive value (marginal increments themselves vary across cycles, as discussed under Convergence below).
    \item \textbf{Starting-Point Effects:} \texttt{S\_generic}, starting from a weaker baseline, achieves the largest cumulative Strict CR gain (+11.60), suggesting that the evolution loop is especially effective at closing the gap for lower-quality initial skills. \texttt{S\_manual}, despite being expert-authored, still benefits substantially (+10.99 Strict, +12.21 Lenient), indicating that automated evolution can surpass human-curated knowledge.
\end{itemize}

\subsubsection{Convergence and Failure Category Analysis}
The per-iteration marginal gains diminish across cycles (e.g., \texttt{S\_manual}: +4.09/+5.55/+1.35 per iteration), indicating convergence toward a performance ceiling. Failure category analysis reveals why: Tool and Style failures decline steadily across iterations ($-$14.5\%/$-$18.2\% and $-$16.4\%/$-$20.9\% respectively), as the optimizer effectively refines tool invocation instructions and response tone. Clarification failures also decline consistently ($-$13.1\%/$-$16.4\%). Knowledge failures, however, plateau after v1 (0\% further reduction in v2)---early iterations resolve the most salient knowledge gaps, but remaining deficiencies prove difficult to close through skill-level optimization alone (we discuss the underlying reasons next), representing the primary driver of convergence.

\subsubsection{Hallucination and the Knowledge Plateau}
\label{sec:hallucination}
The Failure Analyzer's Knowledge dimension explicitly captures factually incorrect or fabricated content (hallucinations); the Knowledge plateau observed above reflects three intertwined limits on automated knowledge refinement. \textit{(i) Reference inconsistency:} different human agents often produce divergent or even contradictory replies to similar tickets---the Optimizer's knowledge search can reconcile some conflicts, but cannot eliminate genuine ambiguity in the historical data. \textit{(ii) Design constraint:} we deliberately restrict the Optimizer from inflating the skill with case-specific facts---the skill body encodes stable, high-frequency domain common sense, the FAQ captures only long-tail items lacking authoritative documentation, and everything else is deferred to the Agent's runtime retrieval rather than baked into the skill. \textit{(iii) Tacit expertise:} expert replies frequently omit the implicit reasoning chains and experiential intuition that practitioners apply to low-frequency, long-tail issues, leaving these patterns difficult to recover from text alone. Together these factors point to an inherent ceiling for purely text-based, fully automated skill optimization: bad-case-driven refinement alone cannot fully eliminate long-tail knowledge errors, however many iterations are run. The human-in-the-loop review (Section~2.1) therefore remains a complementary safeguard for this residual long tail, rather than a substitute for the automated loop.

\subsection{Comparison with Production Legacy System}
We compare v3 against the production legacy system, which combines predefined decision-tree workflows with manually curated expert prompts tuned over an extended period. The skill-equipped agent achieves \textbf{+13.76pp} Strict CR over this legacy system on the same held-out set, confirming that domain-contextualized creation combined with automated self-evolution can surpass mature, human-engineered production systems.
A concrete end-to-end illustration of one evolution cycle---from failure aggregation through diagnosis to the resulting optimization plan---is provided in Appendix~\ref{sec:appendix_case_study}.

% ----------

\section{Related Work}

\subsection{LLM-based Agents and Tool Use}
LLM-based autonomous agents have advanced rapidly, with architectures integrating planning, memory, and tool use into general-purpose problem solvers~\cite{wang2024llmagentsurvey}. ReAct~\cite{react2023} established the dominant paradigm of interleaving reasoning traces with executable actions, while Toolformer~\cite{toolformer2023} and ToolLLM~\cite{qin2024toolllm} demonstrated that LLMs can learn to invoke external APIs at scale. DSPy~\cite{khattab2023dspy} introduced declarative programming abstractions that compile LLM pipelines into self-improving systems. These works operate at the prompt or API-call level, optimizing individual model interactions rather than managing reusable, versionable skill artifacts that encapsulate domain procedures, tools, and knowledge as cohesive packages.

\subsection{Agent Skill Construction and Management}
The concept of agent skills---portable, file-based packages that bundle instructions, tool schemas, and reference materials---was formalized by Anthropic~\cite{anthropicAgentSkills2025} and empirically validated by SkillsBench~\cite{skillsbench2026}, which showed that curated skills improve agent success rates by 16.2pp while self-generated skills provide negligible benefit. Voyager~\cite{wang2023voyager} pioneered automatic skill acquisition through open-ended exploration in Minecraft, building a growing skill library via code-based verification. Recent work has expanded along several axes: SkillX~\cite{skillx2026} constructs hierarchical skill knowledge bases at strategic, functional, and atomic levels; AgentSkillOS~\cite{agentskillos2026} organizes large-scale skill ecosystems via capability trees and DAG-based orchestration; PolySkill~\cite{polyskill2026} introduces polymorphic abstractions enabling cross-domain skill generalization; and AgentFactory~\cite{agentfactory2026} proposes a three-stage lifecycle for progressively accumulating executable sub-agents. TARSE~\cite{tarse2026} explicitly separates reusable skills from episodic experiences for test-time adaptation. Unlike these works, which focus on skill construction, organization, or generalization in open-ended or benchmark settings, SkillForge addresses the distinct challenge of grounding skill creation in enterprise domain knowledge and closing a deployment-feedback-driven self-evolution loop.

\subsection{Self-Evolving Agent Systems}
Self-improvement in LLM agents spans multiple granularities. At the output level, Self-Refine~\cite{madaan2023selfrefine} enables iterative refinement through self-feedback, and Reflexion~\cite{shinn2023reflexion} maintains verbal experience for trial-and-error learning. ExpeL~\cite{zhao2024expel} extracts reusable insights from execution trajectories without parameter updates. At the prompt level, APE~\cite{zhou2023ape}, OPRO~\cite{yang2023opro}, and TextGrad~\cite{yuksekgonul2024textgrad} optimize instructions via search or text-based gradients. At the agent-architecture level, Symbolic Learning~\cite{zhuge2024symbolic} enables post-deployment self-evolution of agent components, ADAS~\cite{hu2024adas} searches over agent designs, and G\"{o}del Agent~\cite{yin2024godel} proposes recursive self-improvement without pre-defined optimization algorithms. Among skill-oriented approaches, MemSkill~\cite{memskill2026} reconstructs memory operations as evolvable skills, EvoSkills~\cite{evoskills2026} introduces co-evolutionary verification between skill generators and verifiers, Steve-Evolving~\cite{steveevolving2026} distills success trajectories into reusable skills and failure trajectories into executable guardrails, and AutoAgent~\cite{autoagent2026} continuously evolves prompt-level cognition with elastic memory. SkillForge differs from these approaches in two key respects: (1) it targets the skill artifact itself---a structured package of instructions, tools, and knowledge---rather than raw prompts or agent architectures; and (2) it operates a structured three-stage pipeline (Failure Analyzer, Skill Diagnostician, Skill Optimizer) that traces deployment failures back to specific skill deficiencies, enabling targeted rather than holistic rewrites.

\subsection{LLM in Enterprise Operations and Customer Support}
LLMs are increasingly deployed in cloud and IT operations. RCACopilot~\cite{chen2024rcacopilot} automates incident root cause analysis at Microsoft, RCAgent~\cite{wang2023rcagent} enables autonomous cloud fault diagnosis with tool-augmented agents, and D-Bot~\cite{zhou2024dbot} applies LLM-driven tree search for database anomaly diagnosis. Broader surveys of LLM-based AIOps highlight the growing adoption of these techniques in failure management~\cite{chen2024aiops}. In customer support, Agent-in-the-Loop~\cite{zhao2025agentloop} implements a continuous improvement flywheel driven by four types of human annotation feedback. SkillForge complements this line of work by automating the feedback loop at the skill level---replacing human annotation with LLM-judged failure analysis and automated skill rewriting---while targeting the specific challenges of cloud technical support where domain knowledge, tool usage, and procedural workflows must be jointly optimized within a single skill package.

\section{Conclusion}
We presented SkillForge, an end-to-end creation--evaluation--refinement framework for domain-specific, self-evolving enterprise agent skills. A Domain-Contextualized Skill Creator grounds initial skill synthesis in historical tickets and domain knowledge, while an automated three-stage pipeline (Failure Analyzer, Skill Diagnostician, Skill Optimizer) continuously diagnoses execution failures and rewrites the skill to eliminate them. Evaluated on five real-world cloud technical support scenarios, \texttt{S\_domain} outperforms the generic baseline by +4.3pp Strict CR and +3.6pp Lenient CR, and the self-evolution loop delivers consistent Strict CR gains of 9--12pp across three iterations regardless of starting point---showing that, in the scenarios studied, automated evolution driven by deployment feedback can surpass manually curated expert knowledge.

\bibliographystyle{unsrt}
\bibliography{references}

\newpage
% ----------------------------------------------------------
\appendix

% Appendix: Domain-Contextualized Skill Creator Details
\section{Domain-Contextualized Skill Creator Details}
\label{sec:appendix_creator}

The Domain-Contextualized Skill Creator addresses the cold-start problem by generating a robust initial skill (\texttt{Skill\_v0}) from historical ticket data and domain documentation. Due to enterprise data privacy requirements, all mining steps use an internal LLM (Qwen3-Max) to process proprietary data. The Creator operates through four sequential stages: Workflow Mining, Tool Mining, Knowledge Extraction, and Skill Synthesis.

\subsection*{Workflow Mining}

Workflow Mining extracts typical solution patterns from historical ticket dialogues, transforming unstructured human-agent conversations into structured resolution traces that capture both explicit actions and implicit expert reasoning.

\textbf{Input.}
The mining operates on individual ticket dialogue segments. Each segment contains the raw conversation between the customer and human agent, along with a coarse category tag (e.g., ``DNS resolution failure'').

\textbf{Process.}
An LLM is prompted to act as a \textit{process mining} engine that reverse-engineers the human agent's thought-action chain from each dialogue. Specifically, the LLM performs the following for every ticket:

\begin{enumerate}
    \item \textbf{Problem Identification}: Extract a detailed, self-contained problem description that enriches the coarse category tag with contextual information from the dialogue, so the description can be understood independently.

    \item \textbf{Resolution Path Reconstruction}: Reconstruct the complete handling workflow by identifying the key phases---clarification (how the agent pinpointed the actual problem), information gathering (what data the agent requested and why), diagnosis/execution (the reasoning chain and actions taken), and solution delivery (the final resolution or escalation).

    \item \textbf{Experience Distillation}: Extract reusable lessons, explicitly labeled as \textit{positive patterns} (effective strategies worth replicating) or \textit{negative warnings} (pitfalls to avoid).

    \item \textbf{Exemplar Response Extraction}: Select high-quality verbatim responses from the human agent that demonstrate effective communication at critical moments---such as de-escalating customer frustration, explaining a complex technical constraint, or guiding the customer through a multi-step operation. Each extracted response is annotated with its usage context (e.g., ``soothing emotion while setting expectations''), enabling the skill to include scenario-specific response templates grounded in proven human phrasing rather than LLM-generated text.
\end{enumerate}

To comply with data governance requirements, the prompt enforces anonymization rules that replace personally identifiable information (names, phone numbers, order IDs, domain names, etc.) with typed placeholders while preserving business semantics.

\textbf{Output.}
Each ticket produces a structured JSON record containing: \texttt{core\_issue} (enriched problem description), \texttt{resolution\_path} (step-by-step narrative), \texttt{accumulated\_experience} (list of labeled lessons), and \texttt{exemplar\_responses} (annotated human agent responses with usage context). These records collectively form the input dataset for subsequent stages.

\textbf{Aggregation.}
After processing all tickets, the individual records are grouped by topic. Statistical analysis identifies high-frequency problem patterns and recurring workflow structures (e.g., Clarify $\rightarrow$ Diagnose $\rightarrow$ Resolve). These aggregated patterns become the backbone of the skill's case-type-specific handling procedures.

\subsection*{Tool Mining}

Tool Mining identifies the tools that human experts frequently invoke during ticket resolution, and extracts their schemas for inclusion in the skill definition.

\textbf{Process.}
Historical tickets contain operation logs that record which internal tools (APIs, diagnostic utilities, configuration interfaces) human agents used during each resolution. The mining step:

\begin{enumerate}
    \item Parses operation logs across all tickets in the target domain to build a tool invocation frequency table.
    \item Applies a frequency threshold to select high-utility tools, filtering out rarely used or deprecated ones.
    \item Extracts the schema for each selected tool (name, description, parameters, return values) from the internal tool registry.
    \item Associates each tool with the scenarios in which it is most commonly used, based on co-occurrence statistics.
\end{enumerate}

\textbf{Output.}
A \texttt{tools.json} file containing the selected tool schemas, along with per-scenario tool usage annotations that inform the skill's workflow steps about when and how to invoke each tool.

\subsection*{Knowledge Extraction}

Knowledge Extraction gathers domain-specific information that the agent needs beyond what is captured in resolution traces---product rules, technical specifications, and troubleshooting references.

\textbf{Process.}
Knowledge is collected from two complementary sources:

\begin{enumerate}
    \item \textbf{Documentation Search}: The system constructs search queries from the task description and identified scenarios, then retrieves relevant articles from the internal knowledge base and official product documentation sites. Retrieved content is filtered for relevance and condensed into concise reference documents.

    \item \textbf{Ticket-Cited References}: Historical tickets often contain links to documentation articles that human agents consulted during resolution. These cited references are collected, deduplicated, and organized by scenario, providing a curated set of authoritative sources validated by actual usage.
\end{enumerate}

\textbf{Output.}
A set of reference documents under \texttt{references/}, each covering a specific knowledge area (e.g., product-specific configuration rules, common error codes and their causes). Core decision-critical knowledge is flagged for inclusion directly in \texttt{SKILL.md}, while detailed background material remains in reference files.

\subsection*{Skill Synthesis}

Skill Synthesis assembles the outputs of the three mining stages into a structured skill package by filling a pre-defined template.

\textbf{Template Structure.}
The skill package follows a fixed directory layout:
\begin{verbatim}
skill_name/
|-- SKILL.md           # Core instructions and workflows
+-- references/
    |-- tools.json     # Tool schemas from Tool Mining
    |-- knowledge_*.md # Domain knowledge from Knowledge Extraction
    +-- ...
\end{verbatim}

The \texttt{SKILL.md} document is organized into the following sections:

\begin{enumerate}
    \item \textbf{Background Knowledge}: Clarifies concepts most commonly misunderstood by customers, sourced from Knowledge Extraction and cross-scenario analysis of mined workflows. This section is intentionally prioritized as the primary determinant of skill quality.

    \item \textbf{Case-Type Triage}: A decision tree derived from the case-type classification in Workflow Mining, enabling the agent to route each incoming request to the appropriate handling procedure based on observable signals.

    \item \textbf{Per-Case-Type Handling} (typically 4--8 case types per skill): Each case type includes an applicability description, a branching workflow reconstructed from mined resolution paths (not a flat procedure), tool invocation guidance from Tool Mining, specific failure causes paired with resolution steps, and an escalation fallback.

    \item \textbf{FAQ}: Covers long-tail issues with frequency below 5\% of the scenario's tickets---too infrequent for a dedicated case type but validated by real occurrence.

    \item \textbf{Reference Index}: Pointers to the reference documents and tool schemas bundled in the skill package.
\end{enumerate}

\textbf{Synthesis Constraints.}
The Creator enforces several quality constraints during synthesis:
\begin{itemize}
    \item All procedures and explanations must trace to actual resolution paths in the mined data, not to general product knowledge.
    \item Handling workflows must encode conditional branching logic rather than presenting linear sequences.
    \item Error symptoms must be paired with specific root causes and resolution steps.
    \item Every case type must include an explicit escalation path for unresolvable cases.
    \item When \texttt{SKILL.md} exceeds 500 lines or 10K characters, detailed content is offloaded to reference files while the main document retains decision logic and workflow steps.
\end{itemize}

% Appendix: Agent Loop Details
\section{Agent Execution Environment Details}
\label{sec:appendix_agent_loop}

SkillForge agents operate within a sandboxed Virtual File System (VFS) that provides complete file operations while ensuring execution safety and traceability.

\subsection{Virtual File System Design}

The VFS is a pure in-memory file system abstraction implemented as a key-value store mapping absolute paths to file nodes. Each node contains metadata (name, type, timestamps, size) and content (for files). The system supports standard operations including read, write, delete, rename, copy, and directory manipulation (mkdir, list, chdir), plus Unix-like utilities (grep, head, tail, find).

All operations return a unified result structure \texttt{\{success, data, error, message\}} without throwing exceptions, simplifying error handling for agents. Path normalization handles relative paths and special directories (\texttt{.}, \texttt{..}) automatically.

\subsection{Integration with Agent Workflow}

At session initialization, the VFS is populated with task-specific files (skill definitions, reference materials). Agents interact through standard tool interfaces (e.g., \texttt{read\_file}, \texttt{write\_file}), with all operations logged in the execution trace for debugging and failure analysis.

The VFS provides complete isolation from the host file system while maintaining familiar file operation semantics, enabling agents to work with complex file structures safely and efficiently.

% Appendix: Failure Analyzer Details
\section{Failure Analyzer Details}
\label{sec:appendix_fa}

The Failure Analyzer performs systematic root cause diagnosis of agent failures through parallel multi-dimensional analysis.

\subsection*{Four-Dimension Analysis Framework}

Each bad case is analyzed concurrently across four dimensions, each producing structured output with severity assessment and issue categorization:

\begin{enumerate}
    \item \textbf{Style Analysis}: Evaluates expression quality (robotic, verbose, cold, inappropriate tone). Style issues only matter when semantic content is correct.

    \item \textbf{Knowledge Analysis}: Identifies knowledge-level problems including missing information, factual errors, contradictions, outdated content, misapplication, or failure to surface existing knowledge.

    \item \textbf{Tool Analysis}: Examines tool invocation behavior for missed calls, wrong tool selection, incorrect parameters, repeated calls, result misinterpretation, or underutilization.

    \item \textbf{Clarification Analysis}: Assesses information gathering strategy appropriateness (over-clarification, under-clarification, wrong clarification focus).
\end{enumerate}

\subsection*{Aggregation Mechanism}

Results from the four dimensions are aggregated through deterministic code logic to produce:

\begin{itemize}
    \item \textbf{failure\_categories}: List of dimensions with detected issues
    \item \textbf{overall\_severity}: Maximum severity across dimensions (high/medium/low/none)
    \item \textbf{overall\_verdict}: fail (high$\geq$1 or medium$\geq$2), marginal (medium=1 or low$\geq$1), or acceptable
    \item \textbf{primary\_category}: Highest severity dimension, with priority order knowledge $>$ tool $>$ clarification $>$ style when tied
\end{itemize}

An optional LLM aggregation step adds natural language summary (\texttt{divergence\_summary}) and actionable diagnostic hints for skill improvement.

\subsection*{Batch Result Aggregation}

After analyzing multiple bad cases, individual results are aggregated by failure category to identify systematic issues. For each category (knowledge, tool, clarification, style), the aggregation computes:

\begin{itemize}
    \item Severity distribution (counts of high/medium/low cases)
    \item Issue type frequencies (e.g., knowledge:missing appears 8 times)
    \item Representative cases (top-k by severity and diversity)
    \item Aggregated diagnostic hints (deduplicated across cases)
\end{itemize}

This category-level aggregation enables the Skill Diagnostician to identify patterns across failures rather than treating each case in isolation, facilitating more effective skill improvements.

% Appendix: LLM-Judge Evaluation Details
\section{LLM-Judge Evaluation Details}
\label{sec:appendix_llm_judge}

The LLM-Judge evaluates agent response quality by comparing against reference responses from human experts. It operates independently from the Failure Analyzer: the Judge determines \textit{whether} a response is acceptable, while the Analyzer diagnoses \textit{why} failures occur. This separation allows the Judge to serve as a proxy for business metrics (e.g., user satisfaction) while the Analyzer focuses on execution trace diagnosis.

The Judge receives four inputs for each evaluation: (1) a global ticket summary containing the customer's core problem, interaction history, and final resolution; (2) the dialogue history between customer and agent up to the current turn; (3) the reference response from a human expert; and (4) the agent's actual response to be evaluated. The Judge first extracts the \textit{core action} from the reference response, filtering out boilerplate (e.g., greetings, closing remarks), and then compares the agent's response against this core action using the three-tier consistency criteria defined in Section~3.1.2. Reference responses are treated as one acceptable solution, not the unique correct answer.

Each evaluation produces structured JSON output: \texttt{\{verdict, ref\_core\_action, actual\_action, reason\}}, where \texttt{verdict} $\in$ \{\texttt{consistent}, \texttt{partial}, \texttt{inconsistent}\}, \texttt{ref\_core\_action} summarizes the substantive action in the reference, \texttt{actual\_action} summarizes the agent's action, and \texttt{reason} provides brief justification for human review. Cases with \texttt{verdict} = \texttt{inconsistent} or \texttt{partial} are forwarded to the Failure Analyzer as bad cases.

\textbf{Human Alignment Validation.} To establish reliability, we validated the Judge against human annotations on a sample subset: each sample was independently labeled by two domain experts, and the Judge achieved over 90\% agreement with the human consensus. This confirms the Judge's reliability as a proxy for expert judgment and justifies its use in the automated bad-case detection loop.

The full evaluation prompt is shown below.

\begin{tcolorbox}[casebox, title=LLM-Judge Evaluation Prompt]
\small
You are a customer service evaluation expert. You need to judge whether an AI agent's ``actual response'' is acceptable in the current dialogue context.

\textbf{[Background]}
You will see four pieces of information:
\begin{enumerate}
\item \textbf{Ticket Summary}: The customer's core problem, interaction history, and final resolution across the entire ticket.
\item \textbf{Dialogue History}: The existing conversation between the customer and agent.
\item \textbf{Reference Response}: The real human agent's response at this point. Note: (a) it may be multiple messages concatenated with line breaks; (b) it may contain boilerplate (e.g., ``Do you have any other questions?'', ``Thank you for your inquiry'') --- extract only the \textit{core substantive content}.
\item \textbf{Actual Response}: The AI agent's response to be evaluated.
\end{enumerate}

\textbf{[Evaluation Criteria]}

The actual response is judged \textbf{Consistent} if \textit{either} condition holds:
\begin{itemize}
\item \textbf{Condition A}: The core action matches the reference (e.g., both request the same information, provide the same link, or guide the same operation).
\item \textbf{Condition B}: The response correctly resolves the customer's problem via an alternative approach that is consistent with the ticket summary's final resolution.
\end{itemize}

The actual response is judged \textbf{Partially Consistent} if:
\begin{itemize}
\item Correct direction but missing key details from the reference (e.g., reference provides specific links/steps, actual response gives only a brief description).
\item Contains correct content with some redundancy but no factual errors or fabrication (no contradiction with the ticket summary).
\item Acceptable action but less optimal than the reference (e.g., customer asked a simple confirmation question; reference answered briefly; actual response is overly verbose).
\end{itemize}

The actual response is judged \textbf{Inconsistent} if:
\begin{itemize}
\item Reference requests information (screenshots, domain names, etc.) but actual response skips this and gives a generic solution.
\item Contains factually incorrect information (contradicts the reference or ticket summary).
\item Completely off-topic.
\item Suggests the customer ``submit a ticket'' or ``contact human support'' (since it \textit{is} the support agent).
\item Reveals its AI identity.
\end{itemize}

\textbf{[Special Notes]}
\begin{itemize}
\item Boilerplate in the reference (greetings, closing remarks, reminders) should not affect the judgment.
\item When the reference is very brief (e.g., ``OK'', ``Please provide your domain name''), the agent only needs to do something simple; over-elaboration is acceptable unless the elaborated content is incorrect.
\item When the customer is expressing gratitude or confirmation and the reference is a polite acknowledgment, any reasonable polite response is Consistent.
\item The reference is not the only correct answer --- it is a reference point. The core criterion is whether the actual response helps the customer.
\end{itemize}

\textbf{[Output Format]}

Output strictly in the following JSON format with no other content:

\texttt{\{``reason'': ``...'', ``verdict'': ``consistent'' | ``partial'' | ``inconsistent'', ``ref\_core\_action'': ``...'', ``actual\_action'': ``...''\}}
\end{tcolorbox}

% Appendix: Skill Diagnostician and Optimizer Details
\section{Skill Diagnostician and Optimizer Details}
\label{sec:appendix_diag_opt}

This appendix describes the Skill Diagnostician and Optimizer, which translate failure analysis results into concrete skill improvements.

\subsection{Skill Diagnostician}

The Diagnostician is implemented as a ReAct agent that traces aggregated failure patterns to specific skill defects.

\subsubsection*{Diagnostic Workflow}

The Diagnostician operates through four stages:

\begin{enumerate}
    \item \textbf{Skill Understanding}: Read and summarize the structure of SKILL.md and reference files
    \item \textbf{Evidence Collection}: Parse aggregated FA results to extract representative cases and diagnostic hints
    \item \textbf{Root Cause Attribution}: For each failure category, map issues to specific SKILL.md locations and classify defect types (missing, insufficient, incorrect)
    \item \textbf{Optimization Plan Generation}: Produce prioritized, actionable modification recommendations with evidence support
\end{enumerate}

\subsubsection*{Attribution Patterns}

Common mappings from FA categories to skill defects include:

\begin{table}[h]
\centering
\caption{Failure Category to Skill Defect Mapping}
\label{tab:attribution_patterns}
\small
\begin{tabular}{@{}p{4cm}p{6cm}@{}}
\toprule
\textbf{FA Category:Type} & \textbf{Likely Skill Defect} \\
\midrule
knowledge:missing & SKILL.md lacks knowledge for this scenario \\
knowledge:not\_surfaced & Knowledge exists but poorly organized \\
tool:missed\_call & No guidance on when to call this tool \\
tool:wrong\_params & Insufficient parameter documentation \\
clarification:over\_asking & Information gathering rules too broad \\
style:verbose & Lacks concise response guidelines \\
\bottomrule
\end{tabular}
\end{table}

\subsubsection*{Diagnostic Report Structure}

The output diagnostic report contains: (1) overview with top issues and category distribution, (2) per-category analysis linking evidence to skill locations, and (3) prioritized optimization plan. Each recommendation specifies the modification location, content changes, whether examples or knowledge search are needed, expected impact, and risk assessment.

\subsection{Skill Optimizer}

The Optimizer executes modifications specified in the diagnostic report, following strict principles to ensure positive evolution.

\subsubsection*{Core Principles}

\begin{enumerate}
    \item \textbf{Minimal Modification}: Only change what's necessary to address diagnosed issues
    \item \textbf{Do No Harm}: Preserve existing correct behaviors through additive changes; never delete working content
    \item \textbf{Evidence-Based}: Every modification must trace to specific FA evidence from bad cases
\end{enumerate}

\subsubsection*{Optimization Workflow}

The Optimizer processes the diagnostic report through these steps:

\begin{enumerate}
    \item Read diagnostic report and identify similar recommendations for merging
    \item Understand original SKILL.md structure to determine appropriate insertion points
    \item Apply modifications by priority, consulting category analysis files for detailed evidence
    \item Add examples when specified (using reference excerpts from FA results)
    \item Perform deduplication check to remove redundant content
    \item Verify changes meet safety criteria (additive, consistent, evidence-backed)
\end{enumerate}

\subsubsection*{Content Placement Strategy}

New content is inserted following the original SKILL.md structure: background knowledge goes in dedicated knowledge sections, tool call rules are embedded in relevant workflow steps, style guidelines appear near response templates, and examples immediately follow related rules. When SKILL.md exceeds size thresholds (500 lines or 10K characters), detailed content is moved to reference files while keeping decision logic in SKILL.md.

\subsubsection*{Knowledge Augmentation}

When the diagnostic report indicates missing knowledge, the Optimizer uses a search tool to retrieve relevant information from the knowledge base. Retrieved content is filtered for relevance, then placed by type following a layered policy: \emph{high-frequency, stable} knowledge is added to the relevant skill section in SKILL.md; long-tail items lacking authoritative documentation are appended to the FAQ; when official documentation is available, the skill cites the source link rather than embedding the content, leaving full retrieval to the Agent at runtime. This policy keeps the skill body lean, avoids inflating SKILL.md with case-specific facts, and is the same layered strategy discussed in the main paper's hallucination analysis.

% \subsection*{Output and Versioning}

% The optimized skill is written to a new version directory (e.g., \texttt{skill-v1/}), preserving the original for comparison and rollback. This versioning enables tracking skill evolution across multiple optimization cycles.

% Appendix: Case Study
\section{Case Study: Skill Self-Evolution}
\label{sec:appendix_case_study}

This appendix presents a concrete case study from the OSS (Object Storage Service) error diagnosis skill, illustrating how the SkillForge pipeline traces agent failures through aggregated analysis to a structured optimization plan.

\subsection*{Skill Overview}

The target skill handles customer tickets related to OSS error diagnosis. The agent's workflow involves: (1) identifying whether the issue is an OSS error (upload failure, access denied, 5xx, etc.); (2) collecting key information (RequestID, bucket name, error messages); (3) invoking diagnostic tools (request log lookup, bucket info query, knowledge search); and (4) providing a precise resolution. The evaluation dataset contains 108 bad cases where the agent's response was judged inconsistent with expert reference responses.

\subsection*{Aggregated Failure Analysis Results}

After the Failure Analyzer processes all 108 bad cases across four dimensions, the aggregated statistics reveal the following distribution:

\begin{table}[H]
\centering
\caption{Failure category distribution across 108 bad cases. Each case may have issues in multiple dimensions.}
\label{tab:case_study_fa_dist}
\small
\begin{tabular}{@{}lrrrl@{}}
\toprule
\textbf{Category} & \textbf{Count} & \textbf{High} & \textbf{Medium} & \textbf{Top Issue Types} \\
\midrule
Style         & 108 & 72 & 36 & verbose (78), robotic (25) \\
Clarification &  97 & 64 & 33 & over\_clarification (59), under\_clarification (31) \\
Knowledge     &  76 & 48 & 28 & missing (48), not\_surfaced (19), incorrect (13) \\
Tool          &  59 & 43 & 16 & missing\_call (34), tool\_missing (21) \\
\bottomrule
\end{tabular}
\end{table}

Three systemic issues emerge from the aggregation: (1) verbose and robotic responses dominate the style dimension; (2) the agent over-clarifies by requesting information the customer has already provided; (3) critical domain knowledge is missing, causing the agent to give incorrect guidance.

\subsection*{Representative Failure Cases}

Below are three representative cases selected from the aggregated results, each illustrating a distinct failure pattern. For each case, we show the divergence summary produced by the Failure Analyzer (explaining how the agent's response diverged from the expert reference) and the automatically generated diagnostic hints.

\begin{tcolorbox}[casebox, title={Case 1: Knowledge Missing --- Signed URL Access to Private Buckets}]
\small
\textbf{Divergence Summary:}
The agent lacked the core knowledge that \textit{a valid signed URL can directly access files in a private OSS bucket}. Instead of confirming accessibility (as the expert reference did), the agent incorrectly demanded error codes and RequestIDs to determine whether the link was accessible, resulting in over-clarification and a verbose response. The expert reference simply confirmed ``yes, it is accessible'' and added that it would incur outbound traffic charges.

\medskip
\textbf{Failure Dimensions:}
\begin{itemize}\setlength\itemsep{0pt}
    \item \textit{Knowledge} (high): Missing knowledge about signed URL access mechanism
    \item \textit{Tool} (high): Did not invoke bucket info query tool to verify bucket ACL; lacked a dedicated signed URL validation tool
    \item \textit{Clarification} (high): Requested unnecessary information when sufficient context (complete signed URL + explicit expiration time) was already provided
    \item \textit{Style} (high): Responded with a lengthy checklist instead of a direct answer
\end{itemize}

\medskip
\textbf{Diagnostic Hints:}
\begin{enumerate}\setlength\itemsep{0pt}
    \item Add the rule ``a valid signed URL can directly access private files'' to the skill's knowledge base
    \item Supplement knowledge about signed URL lifecycle and AccessKey types (temporary vs.\ permanent)
    \item Fix the clarification strategy to avoid requesting error information when a complete signed URL with expiration time is already available
    \item Add tool invocation guidance: when a user provides a signed URL, automatically trigger a bucket ACL query
\end{enumerate}
\end{tcolorbox}

\begin{tcolorbox}[casebox, title={Case 2: Incorrect Knowledge --- Mirror-Based Back-to-Origin Configuration}]
\small
\textbf{Divergence Summary:}
The agent \textit{incorrectly} asserted that OSS mirror back-to-origin configuration cannot contain the \texttt{https://} protocol prefix or \texttt{/*} wildcard, and instructed the customer to modify their (valid) configuration. The expert reference clarified that this format is legitimate and that the 502 error was caused by the origin server not returning a 200/206/404 status code. This knowledge error cascaded into wrong clarification strategy (denying valid configuration before verifying origin behavior), missed tool calls (no bucket config query), and directive-style responses contradicting the facts.

\medskip
\textbf{Failure Dimensions:}
\begin{itemize}\setlength\itemsep{0pt}
    \item \textit{Knowledge} (high): Incorrect understanding of mirror back-to-origin URL format rules
    \item \textit{Tool} (high): Did not invoke bucket info query tool to verify the actual configuration; misinterpreted its own reasoning as ground truth
    \item \textit{Clarification} (high): Directly denied the customer's valid configuration without verification
    \item \textit{Style} (high): Issued authoritative but factually wrong instructions
\end{itemize}

\medskip
\textbf{Diagnostic Hints:}
\begin{enumerate}\setlength\itemsep{0pt}
    \item Correct the skill's description of mirror back-to-origin format rules: protocol prefixes (e.g., \texttt{https://}) and wildcards (\texttt{/*}) are allowed
    \item Add knowledge: when mirror back-to-origin fails, prioritize verifying origin server response status codes (200/206/404) before modifying configuration
    \item Add tool trigger condition: when a user mentions a specific domain or configuration, automatically query actual bucket configuration state
    \item Fix the clarification strategy: do not deny user configuration before confirming origin behavior
\end{enumerate}
\end{tcolorbox}

\begin{tcolorbox}[casebox, title={Case 3: Missing Tool --- Resource Package Billing Diagnosis}]
\small
\textbf{Divergence Summary:}
The agent provided only generic troubleshooting suggestions (checking region mismatch, storage class, etc.) without querying the customer's actual resource package and billing data. The expert reference, in contrast, pinpointed the root cause based on \textit{specific} resource package activation time and billing cycle: ``resource packages can only offset usage incurred \textit{after} the activation date and cannot retroactively offset prior usage.'' The agent listed multiple irrelevant possibilities, resulting in verbose and imprecise guidance. The core gap was the lack of billing data query capability.

\medskip
\textbf{Failure Dimensions:}
\begin{itemize}\setlength\itemsep{0pt}
    \item \textit{Tool} (high): No billing/resource package query tool available; expert reference relied on specific account data (package capacity, activation/expiration dates, billing period) that requires dedicated query capability
    \item \textit{Knowledge} (high): Missing key rule: ``resource packages can only offset usage incurred after activation, not retroactively''
    \item \textit{Clarification} (high): When customer asks about billing anomalies, should prioritize invoking account data query tools rather than giving multi-factor generic answers
    \item \textit{Style} (high): Verbose enumeration of irrelevant possibilities instead of precise diagnosis
\end{itemize}

\medskip
\textbf{Diagnostic Hints:}
\begin{enumerate}\setlength\itemsep{0pt}
    \item Add resource package deduction rule to knowledge base: ``only offsets usage after activation date''
    \item Add resource package and billing query tool to the skill's tool set
    \item Optimize clarification strategy: prioritize tool-based diagnosis over generic multi-factor enumeration for billing inquiries
\end{enumerate}
\end{tcolorbox}

\subsection*{Diagnostic Report and Optimization Plan}

The Skill Diagnostician reads the aggregated failure analysis and the current \texttt{SKILL.md}, then maps failure patterns to specific skill defects. Below is the resulting optimization plan with three prioritized actions.

\begin{tcolorbox}[casebox, title={Optimization Plan --- Priority 1: Fill Critical Knowledge Gaps}]
\small
\textbf{Addresses:} knowledge:missing (48 cases), knowledge:incorrect (13 cases) \\
\textbf{Target locations:} \texttt{SKILL.md} lines 118--121, \texttt{references/common\_errors.md}, \texttt{references/ec\_code\_solutions.md}

\medskip
\textbf{Modifications:}
\begin{enumerate}\setlength\itemsep{0pt}
    \item Add core knowledge: ``a valid signed URL can directly access private files''
    \item Correct mirror back-to-origin format description: explicitly allow protocol prefixes (\texttt{https://}) and wildcards (\texttt{/*})
    \item Add configuration guide for private bucket + CDN back-to-origin with URL rewriting
    \item Add resource package deduction rule: ``packages only offset usage incurred after activation, not retroactively''
\end{enumerate}

\textbf{Requires knowledge search:} Yes (``OSS signed URL private file access'', ``OSS mirror back-to-origin format'', ``OSS resource package deduction rules'') \\
\textbf{Requires examples:} Yes (from expert reference responses) \\
\textbf{Expected impact:} Resolves 48 high-severity knowledge:missing and 13 knowledge:incorrect cases \\
\textbf{Risk:} Low
\end{tcolorbox}

\begin{tcolorbox}[casebox, title={Optimization Plan --- Priority 2: Optimize Clarification Strategy and Response Style}]
\small
\textbf{Addresses:} clarification:over\_clarification (59 cases), style:verbose (78 cases), style:robotic (25 cases) \\
\textbf{Target locations:} \texttt{SKILL.md} lines 40--59 (information gathering), lines 122--139 (response guidelines)

\medskip
\textbf{Modifications:}
\begin{enumerate}\setlength\itemsep{0pt}
    \item Add conditional judgment to information gathering strategy; define ``minimum information set'' concept to avoid requesting already-available data
    \item Add conciseness guidelines: directly answer the core question before expanding
    \item Add natural language and empathy guidance; reduce mechanical checklist-style output
    \item Add tone adjustment rules for negative-emotion scenarios
\end{enumerate}

\textbf{Requires examples:} Yes (concise, empathetic phrasings from expert references) \\
\textbf{Expected impact:} Resolves 59 over\_clarification, 78 verbose, and 25 robotic cases \\
\textbf{Risk:} Low
\end{tcolorbox}

\begin{tcolorbox}[casebox, title={Optimization Plan --- Priority 3: Improve Tool Invocation Logic}]
\small
\textbf{Addresses:} tool:missing\_call (34 cases), tool:tool\_missing (21 cases) \\
\textbf{Target locations:} \texttt{SKILL.md} lines 60--103, \texttt{references/tools.json}, \texttt{references/tool\_usage.md}

\medskip
\textbf{Modifications:}
\begin{enumerate}\setlength\itemsep{0pt}
    \item Specify trigger conditions for the bucket info query tool (e.g., ``when user provides a signed URL, automatically invoke bucket query'')
    \item Add missing tool definitions: resource package/billing query tool, signed URL validation tool
    \item Add tool result interpretation guidance: how to diagnose based on different return values
\end{enumerate}

\textbf{Requires knowledge search:} Yes (``OSS billing query tool'', ``OSS signed URL validation'') \\
\textbf{Expected impact:} Resolves 34 missing\_call and 21 tool\_missing cases \\
\textbf{Risk:} Medium (requires verification that newly added tools are available and accurate)
\end{tcolorbox}

\subsection*{Analysis}

This case study illustrates several key properties of the SkillForge pipeline:

\textbf{Multi-dimensional attribution.} Each failure case is analyzed across four dimensions simultaneously. Case 1, for example, is primarily a knowledge gap, but it cascades into tool, clarification, and style failures---showing that a single root cause can manifest across multiple dimensions. The aggregation step correctly identifies knowledge:missing as the primary category.

\textbf{Evidence-to-location mapping.} The Diagnostician does not produce generic advice. Each optimization item specifies the exact file and line range to modify, the concrete content change, whether knowledge search or examples are needed, and the expected impact in terms of resolved cases. This structured format enables the downstream Skill Optimizer to execute modifications precisely.

\textbf{Prioritized optimization.} The three priorities are ordered by both impact (number of cases resolved) and risk. Knowledge gaps (Priority 1) are addressed first because they are the root cause of cascading failures; style and clarification issues (Priority 2) affect the largest number of cases but are lower risk; tool improvements (Priority 3) have medium risk due to external dependencies on tool availability.

\end{document}